\documentclass[aps, twocolumn, prl, showpacs,preprintnumbers,superscriptaddress]{revtex4-1}

\usepackage{amsmath,amsfonts,amssymb,graphics,graphicx,epsfig,color,times,indentfirst,layout,hyperref,subfigure,multirow,bm}
\usepackage{hyperref}
\usepackage{bbold}
\usepackage{ulem}
\hypersetup{linktocpage,,colorlinks=true,citecolor=red}
\usepackage{soul}

\usepackage{graphicx}
\usepackage{dcolumn}
\usepackage{xcolor}
\usepackage{mathrsfs}
\usepackage{fixmath}
\usepackage[all]{hypcap} % let hyperlinks correctly point to figures rather than their captions; must be loaded after the hyperref package!

\newcommand{\mbd}{\mathbold}

\begin{document}

\title{Quantum phase transition in a two-dimensional Kondo-Heisenberg model: a Schwinger-boson large-$N$ approach}
\author{Jiangfan Wang}
\affiliation{Department of Electrophysics, National Chiao-Tung University, Hsinchu, 300
Taiwan, R.O.C}
\author{Yung-Yeh Chang}
\affiliation{Department of Electrophysics, National Chiao-Tung University, Hsinchu, 300
Taiwan, R.O.C}
\author{Chung-Yu Mou}
\affiliation{Center for Quantum Technology and Department of Physics, National Tsing-Hua University, Hsinchu, 300 Taiwan, R.O.C}
\author{Stefan Kirchner}
\affiliation{Zhejiang Institute of Modern Physics, Department of Physics, Zhejiang University, Hangzhou, P.R.C.}
%\author{Silke Paschen}
%\affiliation{Institute of Solid State Physics, Vienna University of Technology, Wiedner Hauptstr. 8-10, 1040
%Vienna, Austria}
\author{Chung-Hou Chung}
\affiliation{Department of Electrophysics, National Chiao-Tung University, Hsinchu, 300 Taiwan, R.O.C}
\affiliation{Physics division, National Center for Theoretical Sciences, Hsinchu, 300 Taiwan, R.O.C.}
\date{\today}

\begin{abstract}
Strange metal behavior arises in heavy fermion metals close to antiferromagnetic transitions. An increasing amount of experiments indicates a link of such behavior to a Kondo breakdown quantum critical point. To shed light on this intriguing problem, we study the 2D Kondo-Heisenberg model using a dynamical large-$N$ multichannel Schwinger boson approach. We identify and characterize the quantum phase transition from an antiferromagnetically ordered ground state to a Kondo-dominated paramagnetic state, and attribute a jump in certain phase shift to Kondo breakdown. In addition, we calculate transport and thermodynamic quantities and discuss them in the context of the experimental observations in quantum critical heavy fermion systems.
%Strange metal behavior with linear-in-temperature electrical resistivity and logarithmic-in-temperature specific heat coefficient has been widely observed in heavy fermion metals close to antiferromagnetic quantum critical points. The underlying mechanism leading to this behavior constitutes an outstanding and largely unresolved issue. An increasing amount of experiments indicates a link to a Kondo breakdown quantum critical point. To investigate this possibility, we study the 2D Kondo-Heisenberg model via a dynamical large-$N$ multichannel Schwinger boson approach. We identify and characterize the quantum critical point separating the antiferromagnetically ordered from the Kondo-screened heavy Fermi-liquid phases. In addition, we find strange-metal behavior in the finite temperature phase diagram close to the transition, which
%near criticality are well captured within our approach and 
%agrees qualitatively with the experimental results on CeCu$_{6-x}$Au$_x$, YbRh$_2$Si$_2$, and others.
\end{abstract}

\maketitle

``Strange metal  (SM)" behavior, which is characterized by a linear-in-temperature resistivity and logarithmic-in-temperature specific heat coefficient, has been reported in many strongly correlated electron systems close to metallic quantum phase transitions due to competing orders, %magnetic instabilities,
such as in cuprate superconductors \cite{Taillefer-HiTc} and heavy fermion systems \cite{HF-review}. The microscopic origin of this common phenomenology
% as well as the dimensionality associated with the transitions,
 remains an outstanding issue.

 A particularly interesting class of materials with SM features are heavy fermion materials, such as YbRh$_2$Si$_2$ \cite{Trovarelli2000,Custers2010,Custers2003}, CeCu$_{6-x}$Au$_x$ \cite{Lohneysen1994,Schroder2000,Lohneysen1998,LohneysenCeCuAu}, and CeMIn$_5$  with M $=$ Co, Rh \cite{Feng-arpes-CeCoIn, Thompson-CeCoIn-SC, Kirchner-review}, where the systems can be tuned from an antiferromagnetically (AF) ordered phase to a paramagnetic Kondo-screened heavy Fermi liquid (FL) state through a quantum critical point (QCP) \cite{Lohneysen2007}. Low-temperature measurements such as the spin susceptibility of the majority of these materials indicate that the effective degrees of freedom are two-dimensional despite the overall 3D crystal structure \cite{1998CeCuAu2D, Lohneysen2007, Thompson-CeCoIn-SC, Trovarelli-Physica-2000}.
% , many of their low temperature properties exhibit quasi-2D nature via the measurements . 
 Moreover, many experimental findings strongly suggest Kondo breakdown (KB)  \cite{QMSi2001, Si2010} to occur at the AF QCP, most notably the enlargement of the Fermi volume observed via Hall effect measurements \cite{Friedemann2010,Paschen2004},  the $\omega/T$ scaling in the dynamical susceptibility \cite{Schroder2000} and the optical conductivity \cite{Paschen2018arxiv}, and STM measurements \cite{Wirth-STM}. 
 %as well as the vanishing of
%ARPES
% quasiparticle weight \cite{2018-QPW-CeCuAu}. 
 These findings go beyond the standard Hertz-Millis type of spin-density-wave theory \cite{HM-SDW}, critical quasi-particle approaches \cite{CriticalQP}, and phenomenological approaches \cite{Pine-2017-RPP}.
 %Microscopic understanding of these exotic phenomena still remains an outstanding puzzle though some phenomenological approaches have been taken \cite{Senthil2003,Coleman2010,Si2010,Pepin2005}.
\begin{figure}
\centering\includegraphics[width=8.5cm]{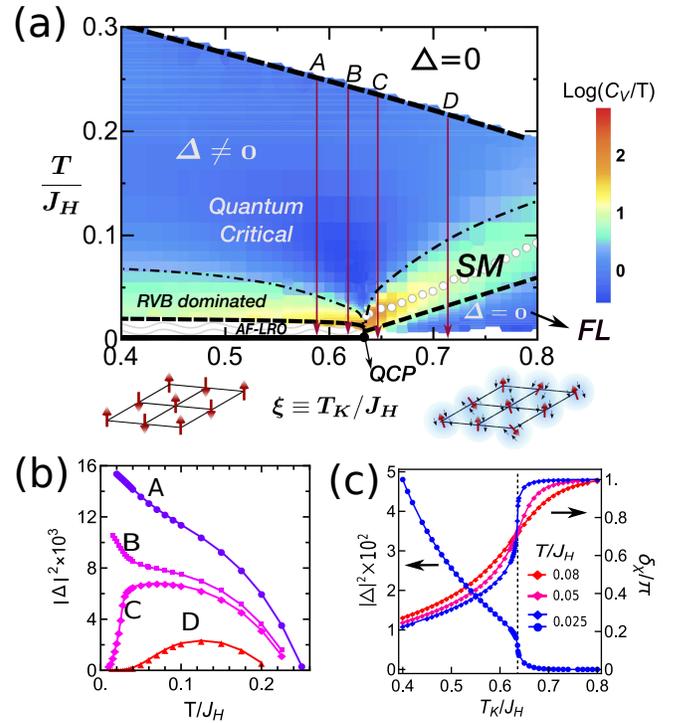}
\caption{(a) Phase diagram mapped out using the specific heat coefficient. The AF and heavy FL ground states are separated by a QCP at $\xi_c \approx 0.635$. Note that the AF-LRO phase only exists at $T=0$ for the 2D KH model (bold solid line). (b) Temperature dependence of the magnetic short range order, $\left| \Delta \right|^{2}$, at different $\xi=T_K/J_H$ marked by $A$ to $D$ in (a). (c) $\left|\Delta\right|^2$ and the holon phase shift $\delta_{\chi}/\pi$ as functions of $\xi$.
}
\label{fig:phase-diag}
\end{figure}

%To account for the nature of the QCP, Doniach's framework for heavy fermion systems where the Kondo effect competes with the Ruderman-Kittel-Kasuya-Yosida (RKKY) interaction in the form of a Kondo-Heisenberg (KH) model is an appropriate starting point. Previous attempts in solving the KH model encountered limitations, e.g., the Dynamical Mean Field Theory (DMFT) and its generalizations for the Kondo and Anderson lattice model \cite{DMFT} have difficulty in describing the competition between the Kondo and RKKY couplings, and its extensions have been devised \cite{QMSi2001, EDMFT, SuncDMFT, BFK}.
%have difficulties in describing the quantum critical antiferromagnetic and Kondo fluctuations. %the Extended DMFT solutions to \cite{ QMSi2001, Zhu2002, Si2003, Si2010} alone or Extended DMFT in combination with numerical renormalization group approach to the Kondo lattice model 
%The extended DMFT \cite{QMSi2001, EDMFT} captures this competition but ignores the fluctuations in Kondo hybridization. Moreover, the underlying Bose-Fermi Kondo model \cite{} is difficult to solve. 
%is restricted to properties at single-impurity level
%The fermionic large-$N$ approach cannot easily describe the magnetically ordered phases \cite{fermionic-Appro}.

To account for the nature of the QCP, Doniach's framework for heavy fermion systems where the Kondo effect competes with the Ruderman-Kittel-Kasuya-Yosida (RKKY) interaction in the form of a Kondo-Heisenberg (KH) model is an appropriate starting point. Kondo breakdown in the KH model has been extensively studied by the EDMFT method \cite{QMSi2001, EDMFT-1, EDMFT-2}, which covers the antiferromagnetically ordered and Kondo-screened heavy Fermi liquid phases as well as the dynamical $\omega/T$ scaling in the quantum critical regime that reflects the dynamical competition between the RKKY and Kondo interactions. Several other techniques have also been used to study this problem \cite{DMFT, DMFT, SuncDMFT, fermionic-Appro, QMSi-BFK-LargeN}, which capture aspects of the Kondo breakdown physics. In this work, we take a dynamical large-$N$ Schwinger boson approach to the 2D KH model, generalized from
%The KH model has been studied previously with various techniques  \cite{DMFT, QMSi2001, EDMFT, SuncDMFT, BFK, fermionic-Appro} but a description of the full temperature-dependent phase diagram, including both the antiferromagnetically ordered and the Kondo-screened heavy Fermi liquid phase as well as the quantum critical/strange metal region, in a single approach remains challenging.
%Different approaches have been used previously to solve the KH model \cite{DMFT, QMSi2001, EDMFT, SuncDMFT, BFK, fermionic-Appro}, but a description of the full phase diagram, including the antiferromagnetically ordered and the Kondo-screened heavy Fermi-liquid phase as well as the quantum critical/strange metal region, in a single approach has not yet been achieved.
%In this work, we offer a different route to this problem by the dynamical large-$N$ Schwinger boson approach to the 2D KH model, generalized from 
the two-impurity Kondo model \cite{Zarand2006}. A simplified version of this approach on a 1D Kondo lattice model was recently studied \cite{Komijani-FM, Komijani-AFM}.
% of Ref.. . 
This method is able to describe both the antiferromagnetic phase through the condensation of bosons as well as
%This approach simultaneously treats antiferromagnetism (via condensing bosons)
 the heavy FL phase. We identify the antiferromagnetic-Kondo breakdown (AF-KB) QCP (see Fig. \ref{fig:phase-diag}(a)) via the holon phase shift linked to the Kondo hybridization and condensation of bosonic spinons. Near this transition, we find a SM region at finite temperatures in between the quantum critical fan and the heavy FL state. We further link the SM behavior to the 2D nature of the critical Kondo and resonating valence bond (RVB) correlations.
 
\begin{figure}
	\centering
	\includegraphics[scale=0.35]{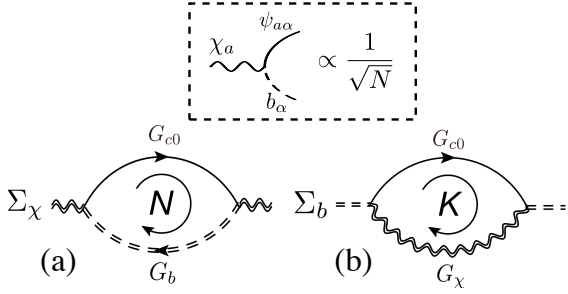}
	\caption{The Feynman diagrams for the self energy of (a) the holon field $\chi$, $\Sigma_\chi$, and (b) the Schwinger boson $b$, $\Sigma_b$. The bare vertex of the Kondo term contributes a factor of  $1/\sqrt{N}$. The double dashed (wavy) line denotes the {\it full} Green's function of the Schwinger boson $b$ ($\chi$) field. The single solid line represents the {\it bare} Green's function of the conduction electrons.}
	\label{fig:feynD}
\end{figure}

\textit{Model- }We start with the KH model $H = H_0 + H_K + H_J$, in which the Kondo coupling $J_K$ ($H_K = J_{K}\sum_{i}\mbd{S}_{i}\cdot \mbd{\sigma}_{i}$) and the AF RKKY interaction $J_H$ ($H_J = J_{H}\sum_{\left\langle i,j\right\rangle }\mbd{S}_{i}\cdot \mbd{S}_{j}$) are treated to be independent of each other for generality. Here, $H_0 = \sum_{\mbd{p}a\alpha}\varepsilon_\mbd{p}\psi^\dagger_{\mbd{p}a\alpha}\psi_{\mbd{p}a\alpha}$ describes the non-interacting conduction electron bath with $\psi^\dagger_{\mbd{p}a\alpha}$ being the creation operator of a conduction electron with quasi-momentum $\mbd{p}$ with channel index $a\in \lbrack 1,K]$ and spin index $\alpha \in \{ -\frac{N-1}{2},-\frac{N-1}{2}+1,\cdots,\frac{N-1}{2}-1,\frac{N-1}{2}\}$. The large-$N$ [Sp($N$)] generalization of $H$ via Schwinger bosons reads \cite{Zarand2006}:
\begin{eqnarray}
H &\rightarrow &H_{0}+\sum_{i}\left[ \frac{1}{\sqrt{N}}\left( b_{i\alpha }^{\dag }\psi
_{ia\alpha }\right) \chi _{ia}+h.c.+\frac{\left\vert \chi _{ia}\right\vert
^{2}}{J_{K}}\right]  \notag \\
&&+\sum_{\left\langle i,j\right\rangle }\left[ sgn(\alpha )b_{i\alpha
}b_{j,-\alpha }\Delta_{ij}+h.c.+\frac{N\left\vert \Delta
_{ij}\right\vert ^{2}}{J_{H}}\right]  \notag \\
&&+\sum_{i}\lambda _{i}\left( b_{i\alpha }^{\dag }b_{i\alpha }-2S\right).
\label{hamil}
\end{eqnarray}
In Eq. (\ref{hamil}), %$H_{0}=\sum t_{ij}\psi _{ia\alpha }^{\dag }\psi _{ja\alpha }$ is the electron hopping term, 
$\psi _{ia\alpha }^\dagger$ is the Fourier component of $\psi _{\mbd{p}a\alpha }^\dagger$ with $i$ being the site index. The Schwinger boson operator $b_{i\alpha }^{\dag }$ creates a spinon of spin $\alpha $. Note that the large-$N$ generalization to the local spin operator $\mbd{S}_i$ at site $i$ can be expressed in terms of either the $N$-flavored bilinear fermions \cite{fermionic-rep} or bosons \cite{Zarand2006, Komijani-FM, Komijani-AFM, boson-rep}. The Kondo hybridization, $\chi _{ia} \propto \sum_{\alpha}\langle \psi_{ia\alpha }^\dagger b_{i \alpha} \rangle$, is a charged, spinless fermionic holon field obtained as a %Grassmann field (also known as holon due to its positive charge) resulting from 
Hubbard-Stratonovich field through decoupling of the original Kondo term. Unlike its bosonic counterpart in the pseudo-fermion representation of spins, though the $\chi_{ia}$ field cannot Bose-condense, i.e.  $\langle\chi_{ia}\rangle  = 0$, % However, by considering its fluctuations, an energy gap develops in both the spinon and holon spectrums, protecting the Kondo singlets 
its fluctuating nature develops an energy gap in both the spinon and holon spectrum which protects the Kondo singlets from being destroyed by thermally excited spinons \cite{Lebanon2007}. As a result, various physical quantities will exhibit Fermi-liquid behavior over a finite temperature window.

The AF Heisenberg $H_J$ %interaction between adjacent local moments is represented in $H$ by the 
is expressed in terms of the Sp($N$)-invariant RVB term, $sgn(\alpha )b_{i\alpha }b_{j,-\alpha }\Delta_{ij}$, where $\Delta _{ij}=-\frac{N}{J_H} \sum_{\alpha} \langle sgn(\alpha)b_{j,-\alpha}^\dagger b_{i \alpha}^\dagger\rangle$ is the AF short-range order (SRO) parameter. The SRO bosonic spin liquid %formed by the Schwinger bosons in our approach allows for a 
is a suitable description of systems with strong magnetic frustration and/or disorder.  %In addition to the short range correlation, the Schwinger boson can develop long-range order (LRO) 
The AF long-range order (LRO) phase is represented in terms of the boson condensate, $\langle b_{i\alpha } \rangle$ \cite{Sachdev1992}. The capability of describing both LRO and SRO is a major advantage of our approach over the fermionic representation. The last term in $H$ reads  %imposes the constraint, 
$n_{b}(i)=\sum_{\alpha } \langle b_{i\alpha }^{\dag }b_{i\alpha } \rangle =2S$, with $2S=K$ corresponding to full Kondo screening \cite{Zarand2006}. The Lagrange multiplier, $\lambda_{i}$, also serves as the temperature-dependent chemical potential of the spinons. Numerically, we treat $\lambda _{i}=\lambda $, $\Delta _{i,i+\hat{x}}=\Delta _{i,i+\hat{y}}=\Delta $ as homogeneous variational $c$-numbers. The ratio $K/N$ is kept constant ($0.2$ in this work \cite{foot-kappa}) as we take the large-$N$ limit.

\textit{Method-} To solve Eq. (\ref{hamil}), we exploit the self-consistent Dyson-Schwinger equations in terms of the fully dressed Green's functions and self energies of various fields. To make further progress, we assume that all the self energies are momentum independent, i.e. $\Sigma \left( i\omega,\mbd{p}\right)\rightarrow \Sigma \left( i\omega \right) $, similar to what is used in the DMFT.  %In this way, the energy dispersions coming from the electron and spinon hopping terms can be integrated out, giving rise to the following
The resulting local Green's functions read
%\begin{eqnarray}
%G_{c}^{-1} &=&i\omega -\varepsilon _{p}-\Sigma _{c}\left(i\omega \right), \notag \\
%G_{\chi }^{-1} &=&-1/J_{K}-\Sigma _{\chi }\left(i\omega \right),\\
%G_{b}^{-1} &=&i\nu -\lambda -\Sigma _{b}\left(i\nu \right) +\frac{4\left\vert \Delta \right\vert ^{2}\xi _{p}^{2}}{i\nu +\lambda +\Sigma_{b}\left(-i\nu \right) },  \notag
%\end{eqnarray}
%and 
\begin{align}
&G_{c0}(i\omega) =\sum_{\mbd{p}} \frac{1}{i\omega -\varepsilon _{\mathbold{p}}},\notag \\
&G_{\chi }\left( i\omega \right) =\left[ -1/J_{K}-\Sigma _{\chi }\left(
i\omega \right) \right] ^{-1},\notag \\
&G_{b}\left( i\nu \right) =\frac{1}{\gamma _{b}\left( i\nu \right) }\frac{2%
}{\pi }E_{K}\left( \frac{16\left\vert \Delta \right\vert ^{2}}{\gamma
_{b}\left( i\nu \right) \gamma _{b}\left( -i\nu \right) }\right) ,
\label{Gb}
\end{align}
and 
\begin{eqnarray}
\Sigma _{\chi }\left( i\omega \right) &=&\sum_{\nu }G_{c0}\left( i\nu
-i\omega \right) G_{b}\left( i\nu \right) ,  \notag \\
\Sigma _{b}\left( i\nu \right) &=&-\kappa \sum_{\omega }G_{c0}\left( i\omega
+i\nu \right) G_{\chi }\left( -i\omega \right).
\label{eq:Sigma}
\end{eqnarray}
In the above equations, $\omega \equiv \pi \left( 2n+1\right) /\beta $ and $\nu \equiv 2\pi n/\beta $ denote the fermionic and bosonic Matsubara frequencies, and $\varepsilon_{\mbd{p}}$ the conduction electron dispersion. In Eq. (\ref{Gb}), $\gamma _{b}\left( z\right) \equiv z-\lambda -\Sigma _{b}\left(z\right)$,   and the complete elliptic integral of the first kind, $E_{K}\left( z\right) \equiv \int_{0}^{\pi /2}1/\sqrt{1-z\sin ^{2}\theta }d\theta $, comes from the 2D momentum integral \cite{SuppleM}. We keep the leading terms in $\Sigma_b$ and $\Sigma_\chi$ of order of unity (Fig. \ref{fig:feynD}), but ignore  %Note that, in Eqs. (\ref{Gb}) and (\ref{eq:Sigma}), we {\color{red}have ignored} 
the vertex corrections and the conduction electron self energy $\Sigma_c$, which are of higer order in $1/N$.% {\color{red}In contrast, both $\Sigma_b$ and $\Sigma_\chi$ are in the order of unity due to the sum over spin or channel indices (see Figure \ref{fig:feynD}).}}

%In addition to the above self-consistent equations, there are
Two constraints for the Green's functions are obtained by minimizing the free energy of the system with respect to $\lambda $ and $\Delta $:
\begin{eqnarray}
\kappa &=&-\int \frac{dz}{\pi }n_{B}\left( z\right) \text{Im}G_{b}\left(z\right) , \nonumber \\
\frac{1}{J_{H}} &=&\int \frac{dz n_{B}\left(z\right)}{2\pi ^{2}\left\vert \Delta \right\vert ^{2}}\text{Im}E_{K}\left[ \frac{16\left\vert \Delta\right\vert ^{2}}{\gamma _{b}\left( z\right) \gamma _{b}\left( -z\right)^{\ast }}\right] .  \label{Constra2}
\end{eqnarray}
Here, $n_{B}\left( z\right) \ $ is the Bose function. % and each argument of $G_{b}$ and $\gamma _{b}$ implicitly contains an infinitesimal imaginary part, $i0^{+}$. 
The unknowns, $G_\chi$, $G_b$, $\Sigma _{\chi }\left( \omega \right) $, $\Sigma _{b}\left( \nu \right) $, $\lambda $, and $\Delta $ are obtained self-consistently through Eqs. (\ref{Gb}) to (\ref{Constra2}).

\begin{figure}
\centering
\includegraphics[width=8.5cm]{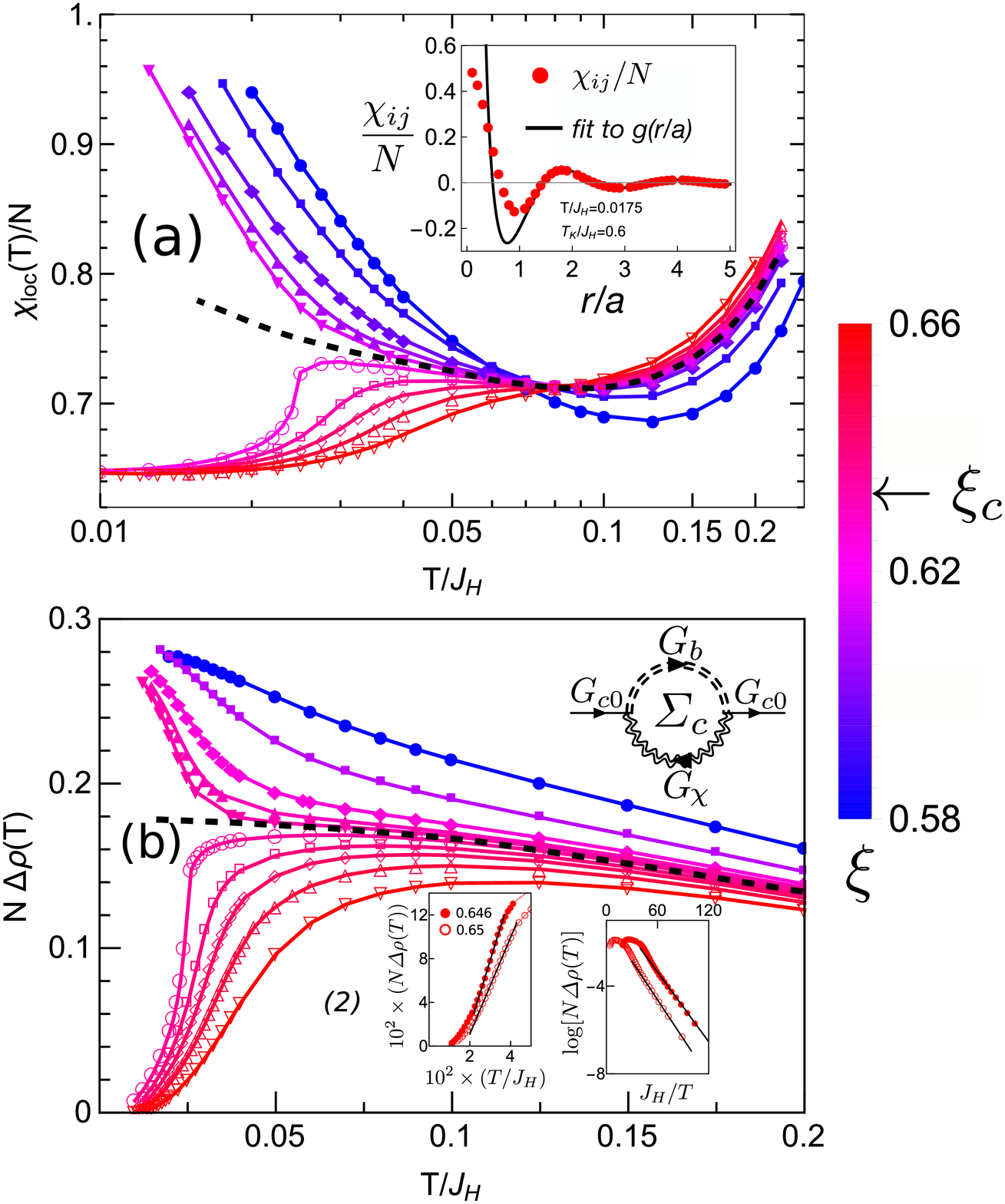}
\caption{(a) Temperature dependence of the local magnetic susceptibility, $\protect\chi _{loc}\left( T\right) /N$ at different values of $\xi$. The schematic dashed line corresponds to $\xi=\xi_c$. Inset: the spatial dependence of the spin susceptibility along the $x$-direction, $\chi_{ij}/N$ (red dots), fitted to $g(r/a)\equiv (a^2 / 5r^2)\cos \left[3.21 r/a\right]$ with $\mbd{r}_i - \mbd{r}_j \equiv (r, \, 0)   $ and $a$ being the lattice constant (solid curve). (b) Electrical resistivity $\rho (T)$ versus $\xi$. Inset: Feynman diagram of the self energy of a conduction electron, $\Sigma_c$. The resistivity behaves as $\rho (T) \sim \exp (-J_H/T)$ ($\rho(T)\sim T$) in the lower (higher) temperature regime. The dashed lines in (a) and (b) schematically show the extrapolations of $\chi_{loc}$ and $\rho(T)$ at the QCP. 
}
\label{fig:scaling}
\end{figure}

\textit{Global phase diagram-} Our main results are summarized in Fig. \ref{fig:phase-diag}(a): a phase diagram in terms of the tuning parameter $\xi \equiv T_K/J_H$ ($T_{K}=De^{-D/J_{K}}$ is the bare Kondo temperature with $D$ being the half bandwidth of the conduction electrons) and the temperature $T/J_H$ is mapped out via the specific heat coefficient $C_V/T$. At $T=0$, a QCP at $\xi =\xi_c$, separating the AF-LRO phase at small $\xi$
% where rotational symmetry breaks, 
 and the paramagnetic FL ground state at large $\xi$
% , where all the impurity spins are completely screened
 is clearly identified via the low-temperature evolution of various quantities. The AF-LRO phase can be inferred from the spinon spectral function showing a non-zero weight of massless mode (or condensation of bosons $\left\langle b\right\rangle \neq 0$), while the heavy FL phase is identified through the linear-in-$T$ entropy and the holon phase shift, defined as
\begin{align}
	\frac{\delta_\chi}{\pi}=-\text{Im}\ln \left[ 1+J_{K}\Sigma _{\chi }\left( i0^{+}\right) \right].
\end{align} 
Within this phase, the magnetic SRO vanishes, and both the spinons and holons develop gaps in their spectral functions. 
%Figure \ref{fig:phase-diag}(b) and (c) show the temperature dependence of $|\Delta|^2$, and the $\xi$-dependence of both $|\Delta|^2$ and $\delta_\chi/\pi$, respecively. 
The abrupt jump of holon phase shift $\delta_\chi/\pi$ (Fig. \ref{fig:phase-diag}(c)) and the vanishing of $|\Delta|^2$ as $T\rightarrow 0$ for $\xi > \xi_c$ (Fig. \ref{fig:phase-diag}(b)) indicate a reconstruction of the Fermi surface at the QCP, which is consistent with the Kondo breakdown observed in various heavy fermion compounds including YbRh$_2$Si$_2$ \cite{Paschen2004, Friedemann2010} and CeCu$_{6-x}$Au$_x$ \cite{LohneysenCeCuAu, Schroder2000}. In addition, the AF-SRO is completely suppressed in the heavy FL phase, different from the previous studies for the two-impurity and the 1D Kondo lattice model \cite{Zarand2006,Komijani-FM, Komijani-AFM}. 
%At non-zero temperatures, the region in parameter space with non-zero RVB strength $(|\Delta|\neq 0)$ extends to the quantum critical region. As a result,
We find an AF-LRO phase to persist at finite temperatures [Fig. \ref{fig:scaling}(a)]. However, this finite temperature LRO phase is ruled out by the Mermin-Wagner theorem \cite{MWreview} and thus is an artifact of our approach \cite{footnote3}. Above the LRO phase, an RVB-dominated region, where $|\Delta|$ is significantly enhanced, is found. Near the QCP, the SM behavior, characterized by a logarithmic-in-$T$ specific heat coefficient and linear-in-$T$ resistivity, is clearly seen within a narrow region on the Kondo side (marked as ``SM'' in Fig. \ref{fig:phase-diag}(a)). As further explained in the Discussion section, we believe that the SM region being rather narrow and tilted towards the Kondo-screened side is an artifact of our approach. 
%Nevertheless we point out that its presence is a key new result.
% We believe  this distinct shape of the SM region from the conventional quantum critical fan is also an artifact of our approach. See discussion.

%Although the finite temperature LRO phase is , our approach appears to  so that the  appears to show the emergence of long-range order as indicated by the inset of Fig. \ref{fig:scaling}(a) .

 %In addition, we found that the LRO phase expands upon increasing $K/N $, consistent with the zero temperature result of the 2D Heisenberg model \cite{Read1991}.

\begin{figure}
\centering
\includegraphics[width=8.2cm]{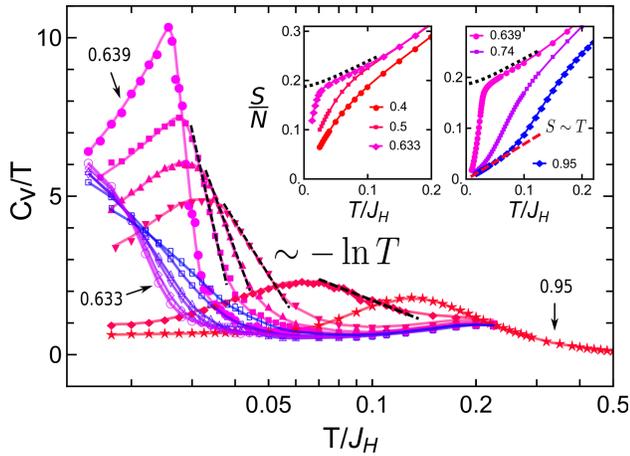}
\caption{
(a) Temperature dependence of $C_V/T$ at different values of $\xi$ [from $\xi=0.61$ (blue) to $\xi=0.95$ (red)]. A $\ln T$ NFL behavior is identified (dashed lines). Inset: Entropy per spin as functions of temperature at $\xi<\xi_c$ (left) and $\xi > \xi_c$ (right), respectively. At $T = 0$, there is a residue entropy at the QCP, as indicated by the black dotted curves. 
%A linear-in-$T$ dependence in $S/N$ at low temperatures is observed (red dashed line), showing an evidence of the Fermi liquid phase.
 }
\label{entropy}
\end{figure}

\textit{Entropy and specific heat coefficient-} Figure \ref{entropy} shows the temperature dependence of the entropy, $S$ \cite{Coleman2005b}, and the specific heat coefficient, $\gamma \equiv C_{V}/T=\partial S/\partial T$ at different values of $\xi$.
% {\color{red}by} using the formula of Ref. . 
% The same QCP where $\xi _{c}=0.634$, is found in the color-coded plot of $\ln \left(C_{v}/T\right) $ {\color{red}(see Figure \ref{entropy}(a))}. 
 For $\xi>\xi_c$, the plateau at low temperatures in the specific heat coefficient associated with the linear-in-$T$ entropy reveals the formation of the FL phase. As $\xi$ approaches $\xi_c$ from the Kondo side, the temperature range of the plateau shrinks monotonically and vanishes at the QCP. Above the FL region, the specific heat coefficient increases till it reaches a ``Schottky" peak \cite{foot-Schottky}. Above the Schottky peak, the specific heat coefficient decreases logarithmically with increasing temperatures, with a scaling form $b(\xi) C_V/T \sim -\alpha(\xi)\ln(T/T^\ast)$ with $b$ and $\alpha \sim |\xi - \xi_c|^{-0.49}$ being two non-universal factors \cite{SuppleM}. Similar strange metal behavior in $C_V/T$ has been observed in YbRh$_2$Si$_2$ \cite{Custers2003} and CeCu$_{6-x}$Au$_x$ \cite{Lohneysen1998}. This logarithmic singularity in the specific heat coefficient is a characteristic of 2D bosonic critical fluctuations, reminiscent of similar features recently obtained via a fermionic large-$N$ approach to Ge-substituted YbRh$_2$Si$_2$ and CeMIn$_5$ \cite{2018SMChang}. Note that, the peak temperatures (open circles in Fig. \ref{fig:phase-diag}(a)) display a weakly first-order jump at the QCP, similar to that in Ref. \cite{Heisenberg-SWB}  (see Discussion section for this point).

 % Figures \ref{entropy}(b) and (c) show the temperature dependence of $C_{v}/T$ and the entropy at different values of $\xi $, respectively.
 % The inset of Figure \ref{entropy}(b) shows a log-linear plot of $C_v/T$ at two different $\xi$.  
% On the Kondo side, the entropy becomes almost linear at low enough temperatures, indicating the formation of Fermi liquid state.
%Our results for specific heat share a qualitative resemblance to that observed in YbRh$_{2}$Si$_{2}$.

\textit{Magnetic susceptibility- }The static local (momentum integrated) magnetic susceptibility $\chi_{loc} (T)$ as a function of temperature defined as
\begin{equation}
\frac{\chi _{loc}\left( T\right) }{N}=2\int \frac{dz}{\pi }n_{B}\left(z\right) G_{b}^{\prime }\left( z\right) G_{b}^{\prime \prime }\left(z\right) 
\end{equation}
 is plotted in Fig. \ref{fig:scaling}(a) at different values of $\xi $ near the QCP. It shows two distinct  properties corresponding to the two regimes $\xi>\xi_c$ and $\xi<\xi_c$ as $T/J_H \rightarrow 0$. In the Kondo limit, the susceptibility acquires a crossover from a saturated Pauli susceptibility  at low temperatures, where all the local spins are fully screened, to a typical spin liquid behavior at relatively high temperatures \cite{Auerbach-Winterfeldt}. Meanwhile, both the local and the uniform susceptibility exhibit a Curie's law at temperatures above the spin liquid region \cite{SuppleM}. The  quantum critical fan is identified through the power-law scaling of the local susceptibility \cite{SuppleM}.
  %$\chi_{loc} \sim T^{-0.035+\alpha}$ with $\alpha \propto \xi-\xi_c$ (the dashed line in Fig. \ref{fig:scaling}(a) corresponds to $\xi=\xi_c$). 
  Further experiments on inelastic neutron scattering or Knight shift measurements are needed to confirm this critical behavior.

\textit{Electrical resistivity-} Figure \ref{fig:scaling}(b) show the electrical resistivity obtained through the the Boltzmann formula \cite{HewsonBook},
\begin{equation}
\rho^{-1}(T)  =-\frac{ne^{2}}{m}\int \tau \left( \omega \right)
\frac{\partial n_{F} (\omega)}{\partial \omega }d\omega ,
\end{equation}
where $n_F (x) \equiv [\exp (x/T)+1]^{-1}$ denotes the Fermi function and $\tau ^{-1} \left( \omega \right) =-2 \Sigma _{c}^{\prime \prime}\left( \omega \right) $ is the scattering rate of the conduction electrons [of order of $O(1/N)$, see inset of Fig. \ref{fig:scaling}(b)]. In the Kondo limit, we observe a typical NFL linear-in-$T$ dependence of the electrical resistivity $\rho (T)= a + bT$ with $a$ and $b$ being constants [inset of Fig. \ref{fig:scaling}(b)]. The $T$-linear resistivity is more clearly featured as the system approaches the QCP, suggesting its link to the critical Kondo fluctuations \cite{2018SMChang}, similar to what is seen in pure YbRh$_2$Si$_2$ \cite{Trovarelli2000, Custers2003}, Ge-substituted YbRh$_2$Si$_2$ \cite{Custers2010}, and CeMIn$_5$ \cite{Feng-arpes-CeCoIn, Thompson-CeCoIn-SC}. On the AF side, insulating behavior is observed at low temperatures.
%The overall line shape and crossovers of $\rho(T)$ show much resemblance with that seen in  CeCu$_{6-x}$Au$_x$ \cite{Lohneysen1998}. 
Note that, to the leading order in $N$, the expected $T^2$ Fermi liquid behavior is replaced by an exponential decay [inset of Fig. \ref{fig:scaling}(b)], due to the finite holon and spinon gaps \cite{Lebanon2007}. 

\textit{$\omega/T$ scaling-} The $\omega/T$ scaling has initially been observed in the dynamical spin response \cite{Schroder2000}, and connected to the critical Kondo Breakdown \cite{QMSi-BFK-LargeN}. More recently, dynamical scaling has also been demonstrated in the charge response \cite{Pixley-KB} and observed in optical conductivity measurements on YbRh$_2$Si$_2$ \cite{Paschen2018arxiv}, which is thus believed to be induced by the emergent critical charge fluctuations at the KB QCP. This observation is supported by the $\omega/T$ scaling of the critical valence fluctuations in $G_\chi$ at the KB QCP,
%The $\omega/T$ scaling recently observed in optical conductivity measurements on YbRh$_2$Si$_2$ \cite{Paschen2018arxiv} is believed to be induced by the emergent critical charge fluctuations at the KB QCP. 
%In our model, this soft charge fluctuation is described by the fermionic holon field, $\chi$, as 
%This observation is supported by the $\omega/T$-scaling of $G^{\prime\prime}_\chi$ at the QCP, 
 i.e. $ G^{\prime\prime}_{\chi}(\omega, T)= -T^{0.56}f(\omega/T)$ with $f(\omega/T)$ being an universal function \cite{SuppleM}, consistent with the linear-in-$T$ behavior of $\Sigma^{\prime\prime}_c (\omega = 0 , \, T)$, i.e. $\Sigma^{\prime\prime}_c (\omega = 0 , \, T) \sim T$. %This provides another advantage of our approach over other methods like DMFT.

{\it Discussions-} Before concluding, we would like to make some remarks. Firstly, at non-zero temperatures, the region in the parameter space with non-zero RVB strength $(|\Delta|\neq 0)$ extends to the quantum critical region. As a result, the SM behavior can only be observed within a narrow region on the
Kondo side of the QCP.
%, and the range gets smaller as approaching QCP,
%the SM state only occupies a small region in our phase diagram and the temperature range with  SM behavior becomes narrower as approaches the QCP, 
%which is in stark contrast to the experimental findings.
 We attribute this feature to the bosonic representation of impurity spins where the holon field $\chi$ cannot get Bose-condensed, giving rise to an overestimated RVB mean field $\Delta$. We expect that introducing a combined boson-fermion supersymmetric representation of spins \cite{ColemanSUSY2016}, where $\chi$ and $\Delta$ are treated on equal footing, may remedy this artifact. 
%However, two scenarios can be expected in this case: either the mean-field $\Delta$ and $\chi$ vanish at the same QCP, or they coexist over a finite range in $T_K/J_H$. In either case, the SM behavior near the QCP is expected to arise as a result of the competitions between the two fluctuating order parameters. 
Secondly, weakly first-order transitions usually exist in the  Schwinger boson mean-field theories due to the attractive quartic term of order $\mathcal{O}(\Delta^4)$. This term can be cancelled via including a small repulsive biquadratic term $J^\prime_H (\mbd{S}_i \cdot \mbd{S_j})^2$ with $J^\prime_H \ll J_H $ in the Hamiltonian and thus the second-order phase transitions can be restored \cite{Komijani-FM}. Thirdly, the finite-temperature AF-LRO phase we find here is likely due to an overestimated Kondo-induced long-range RKKY interaction \cite{footnote3}, as manifested by the long-range correlated position-dependent magnetic susceptibility [inset of Fig. \ref{fig:scaling}(a)].
%, the AF long-range order at finite temperatures can be stablized.
 Nonethelss, this long-range order will be stablized in the form of 3D long-range order in real materials due to weak inter-layer coupling and the calculated 2D long-range order will emerge as quasi-long-range order in the corresponding renormalized classical regime \cite{Nelson1989}.

\textit{Conclusions- }We have explored the quantum phase transition and the quantum criticality of heavy fermion compounds based on the antiferromagnetic Kondo-Heisenberg model on a two-dimensional square lattice via the large-$N$ [Sp($N$)] multichannel Schwinger boson approach. The global phase diagram and the behavior of various physical observables therein show a close resemblance to YbRh$_2$Si$_2$ \cite{Trovarelli2000, Custers2010,Custers2003, Paschen2004, Friedemann2010, Paschen2018arxiv} and CeCu$_{6-x}$Au$_x$ \cite{Schroder2000, Lohneysen1994,Lohneysen1998, LohneysenCeCuAu}: At $T=0$, an antiferromagnetic-Kondo breakdown quantum critical point, which separates the antiferromagnetic long-range order phase from the Kondo-screened heavy Fermi-liquid phase, is identified via the low temperature evolution of several physical observables. The strange metal state with $T$-linear electrical resistivity and a logarithmic divergence in the specific heat coefficient is also observed. The universal critical scaling of the local susceptibility further underpins the existence of a quantum critical region. Further studies to generalize our work to the supersymmetric dynamical large-$N$ approach are needed.
%An extension of this work is to benchmark our model with DMFT to see whether the multichannel Schwinger boson approach is more economic in describing heavy fermion systems.

  % a) the mean-field  $|\Delta|$ and $\left\langle\chi \right\rangle$ each occupies one side of the QCP, leaving a large in-between quantum critical region, and b) the two mean-field phases have an overlap region covering the QCP, naturally giving rise to a superconducting dome and possibly a strange metal region above it. In both scenarios, the strange metal could be a result of the competition between the two fluctuating order parameters.}

The authors acknowledge discussions with P. Coleman, Y. Komijani, G. Kotliar, S. Paschen, and Q. Si. This work is supported by the MOST (104-2112-M-009-004-MY3), the MOE-ATU program and the NCTS of Taiwan (C.H.C.); Center for Quantum Technology from the Featured Areas Research Center Program within the framework of the Higher Education Sprout Project by the Ministry of Education (MOE) in Taiwan (C.Y.M.);  National Science Foundation of China (11774307) and the National Key R$\&$D Program of MOST of China (2016YFA0300200 and 2017YFA0303100) (S.K.).
%  the Austrian Science Fund (FWF W1243 and P29279-N27) (S.P.).


\begin{thebibliography}{36}
\bibitem{footnote3}Our approach does obey the Mermin-Wagner theorem in the 2D Heisenberg model, corresponding to $J_K=0$ in our model.
\bibitem{foot-Schottky}The occurrence of Schottky peak in the specific heat coefficient is a consequence of the evolution of the changes of entropy with temperatures.
\end{thebibliography}
\end{document}